\documentclass[preprint]{aastex}

\shortauthors{LOCKMAN \& CONDON}

\begin{document}
\title{The Spitzer Space Telescope First-Look Survey: \\ Neutral Hydrogen
Emission}

\author{Felix J. Lockman}
\affil{National Radio Astronomy Observatory\altaffilmark{1},
  P.O. Box 2, Green Bank, WV, 24944} 
\email{jlockman@nrao.edu}  \and 
\author{J.~J. Condon} \affil{National Radio Astronomy Observatory, 
520 Edgemont Rd., Charlottesville VA 22903} 
\email{jcondon@nrao.edu}
\altaffiltext{1}{The National Radio Astronomy Observatory is operated
by Associated Universities, Inc., under a cooperative agreement with the
National Science Foundation.}

\begin{abstract} 
The Spitzer Space Telescope (formerly SIRTF)
extragalactic  First-Look Survey covered
about 5\,deg$^2$ centered on J2000 $\alpha = 17^{\rm h} 18^{\rm m}$,
$\delta = 59^\circ 30'$ in order to characterize the infrared sky with
high sensitivity.  We used the 100-m Green Bank Telescope to image the
21\,cm Galactic H\,I emission over a $3\arcdeg \times
3\arcdeg$ square covering this position 
with an effective  angular resolution of $9\farcm8$ and a 
velocity resolution of 0.62\,km\,s$^{-1}$.  In the central square
degree of the image the average column density is $N_{\rm HI} = 2.5
\times 10^{20}$\,cm$^{-2}$ with an rms fluctuation of $0.3 \times
10^{20}$\,cm$^{-2}$.  The Galactic H\,I in this region has a very
interesting structure.  There is a high-velocity cloud, several
intermediate-velocity clouds (one of which is probably part of the
Draco nebula), and narrow-line low velocity filaments.  
The H\,I emission shows a strong and detailed correlation with dust.
Except for the high-velocity cloud, all features in the $N_{\rm HI}$ 
map have counterparts in an $E(B-V)$ map derived from infrared data.  
Relatively high $E(B-V)/N_{\rm HI}$ ratios in some directions
suggest the presence of molecular gas.  The best diagnostic of
such regions is the peak H\,I line brightness temperature, not the
total $N_{\rm HI}$: directions where $T_{\rm b} > 12$\,K have
$E(B-V)/N_{\rm HI}$ significantly above the average value. 
The data corrected for stray 
radiation have been released via the Web.
\end{abstract}
 
\keywords{Galaxy: structure --- infrared: ISM --- radio lines: ISM ---
surveys}

\section{Introduction}

The Space Infrared Telescope Facility (SIRTF) was launched in 2003 and
renamed the Spitzer Space Telescope.  Following the in-orbit checkout,
it began the First-Look Survey (FLS) of selected areas to characterize
the far-infrared (FIR) sky two orders of magnitude deeper than was
reached by previous instruments (see {\tt
http://ssc.spitzer.caltech.edu/fls/}).  The extragalactic portion of
the FLS covers about $5\,{\rm deg^2}$ centered on J2000 $\alpha = 17^{\rm h}
18^{\rm m}$, $\delta = +59^\circ 30'$.  Most of the extragalactic FIR
sources are expected to have radio continuum counterparts obeying the
FIR/radio correlation, so this region has been imaged by the VLA with
matching sensitivity at 1.4\,GHz \citep{con03}.  Because there is a
good correlation between Galactic FIR emission and the column density
of H\,I at high Galactic latitudes \citep{boulanger88, boulanger96,
schlegel, hauser}, high-quality 21cm H\,I data from this region can be
used to estimate foreground Galactic cirrus emission which might
confuse extragalactic studies.  The H\,I data are also useful for
studying properties of Galactic dust seen in the FIR (e.g.,
Boulanger et al.~2001).

We have observed the area of the FLS survey in the 21\,cm line of H\,I
 with the Green Bank Telescope (GBT).
The data give accurate values of the neutral hydrogen column density
$N_{\rm HI}$ throughout the field and reveal interesting interstellar
structures, most of which are visible in existing FIR data.

\section{Observations and Data Reduction}

We used the Robert C. Byrd Green Bank Telescope (GBT)
\citep{fjlspie98,jewellspie} to measure the 21\,cm H\,I emission over the 
FLS field six times during three observing sessions in 2002, 2003, and
2004.  The GBT has a half-power beam width of $9\farcm2$ in the 21\,cm
line at 1.420~GHz.  The total system temperature toward the FLS field
was $\leq20$\,K, and the receiver recorded both circular polarizations.
For this experiment the telescope was scanned in right ascension at a
constant declination, then the declination was stepped by $3'$ and the
scan direction was reversed.  The resulting images cover an area of
$3\arcdeg \times 3\arcdeg$ in right ascension and declination
centered on J2000 $\alpha = 17^{\rm h} 18^{\rm m}$,
$\delta = +59\arcdeg 30\arcmin$ ($\ell = 88\fdg32$,
$b = +34\fdg89$).  Within this area H\,I spectra were measured
every 3\arcmin\ in both coordinates.  The integration time was 3\,s
per pixel for each of the six maps, and data were taken 
by frequency switching out of the band.  The spectra cover
about 520\,km\,s$^{-1}$ centered at $V_{\rm LSR} = -50$\,km\,s$^{-1}$ with
a channel spacing of 0.52\,km\, s$^{-1}$ and an effective velocity
resolution of 0.62\,km\,s$^{-1}$.

The spectral brightness temperature scale was determined from
laboratory measurements of the noise diode and checked with frequent
observations of the standard regions S6 and S8 \citep{williams,kmr82}.
We used aips++ to regrid and average the spectra  into 
a single data cube with a pixel size of $1\farcm5$.
  Third-degree polynomials 
were fit to line-free regions of the spectra and subtracted, and 
a  correction for stray radiation was applied as
described below, yielding a final data cube with the properties given
in Table 1.  The regridding broadened the effective angular resolution
to a FWHM of $9\farcm8$.

\subsection{Stray Radiation}

Ideally, all signals received by a radio telescope would come through
the main antenna beam. Real telescopes, however, always have sidelobes.
These sidelobes cause special problems for observations of Galactic H\,I
because there is Galactic 21\,cm H\,I emission from every
direction on the sky.  All telescope sidelobes not falling on the 
ground thus contribute 
``stray'' radiation which must be subtracted if accurate H\,I spectra
are to be obtained.  Unlike all other large antennas, the GBT has an
unblocked optical path and should have a minimum of stray H\,I
radiation.  However, the feed horn currently used for 21\,cm
observations is part of a general purpose L-band system covering 1.15\,GHz --
1.73\,GHz from the Gregorian focus.  Because of structural limitations
on its size and weight, this feed over-illuminates the 8\,m diameter
secondary reflector at the 1.420\,GHz frequency of the 21\,cm line,
creating a broad and diffuse forward spillover lobe which contains
about 4\% of the telescope's response (Norrod \& Srikanth
1996; S.~Srikanth, private communication).  Efforts are underway to
model and measure this sidelobe.  Its importance for observations of
the FLS field is illustrated by Figure 1, which shows the striking
difference between two GBT H\,I spectra of the same sky area  
observed at different local sidereal times (LSTs).  The differences between
the spectra are due entirely to changes in the stray radiation.  In
Figure 2, the the two ovals show the GBT subreflector in Galactic
coordinates as seen from the focal point at the times the two
observations were made.  When the upper rim of the subreflector lies
at low latitude, e.g., at 23$^{\rm h}$ LST, the spillover lobe picks
up the very bright H\,I emission from the Galactic plane, whereas at
10$^{\rm h}$ LST the subreflector rim lies primarily on faint H\,I at
high Galactic latitude.  The general characteristics of the profiles
in Figure 1 suggest that this understanding of the origin of the
difference between the spectra is correct.

The best way to compensate for stray radiation entering a
telescope's sidelobes is to determine the telescope's response in all
directions, estimate the contribution from H\,I in directions away from the
main beam, and subtract that from the observed spectra.  This
method has been used successfully to correct data from several
instruments \citep{kmr, LDstray, higgs}, but it is quite laborious
and requires knowledge of a telescope's sidelobes in directions where
they are very weak.  Moreover, it is not certain that developing this
technique for the GBT will be worth the effort because,
unlike conventional reflectors whose sidelobes are caused by aperture blockage
which can never be eliminated,  the strongest
sidelobes of the GBT are consequences of the specific feed
and are not at all fundamental. 

For the immediate purposes of this work, we estimated the stray
component in the FLS field by ``bootstrapping'' the GBT data to the
Leiden-Dwingeloo survey data in the general manner described by
\citet{ljm} and \citet{fjl2003}.  This method assumes that the stray
component of a 21\,cm spectrum, which typically arises from 
broad sidelobes, does not change significantly with changes in
main-beam pointing of a few degrees, and is also approximately constant
during the few hours it takes to make one map. 
 Each GBT map of the FLS field was made over 3 hours, so it
is plausible that over this period the stray radiation spectrum can be
approximated by a single average profile.  That profile can be
determined by convolving the GBT H\,I image to the angular resolution
of the Leiden-Dwingeloo (hereafter LD) survey, which was made by a
25\,m telescope having 36\arcmin\ resolution \citep{LDsurvey} and was
corrected for stray radiation using an all-sky model of its response
\citep{LDstray}.   Any difference between the GBT data
convolved to 36\arcmin\ resolution, and the LD survey spectra at the
same position, can be attributed to stray radiation in the GBT data.

We applied this technique to the GBT observations of the FLS.  The GBT
data were convolved to the 36\arcmin\
resolution of the LD survey and compared with LD observations in the
central area of the FLS field using a revised version of the LD survey 
with an improved stray-radiation correction kindly 
supplied by P. Kalberla.  The difference spectra  have the
expected characteristics of stray radiation given the location of the GBT
spillover lobe on the sky at the times of the observations.  
There are two exceptions, however: (1) the high-velocity cloud at $V_{\rm LSR}
\approx -190$\,km\,s$^{-1}$ and (2) the narrow, bright, spectral feature
near $V_{\rm LSR}=0$ which appears in many directions over the FLS
field.  For both features the naive ``bootstrapping'' procedure gives
inconsistent, implausibly large, and sometimes unphysically negative
corrections to the GBT data.  We believe that these problems are
caused by limitations in the LD survey and its stray-radiation
correction.  The LD survey used its own data as the H\,I sky for
determining stray-radiation corrections in an iterative procedure, but
that H\,I sky was undersampled, being measured only every $0\fdg5$.
Thus real features near their main beam that have angular structure or
velocity gradients on scales $\lesssim 0\fdg5$ will be aliased into an
erroneous stray-radiation correction.  The potential for this sort 
of systematic effect  was noted by
the LD survey group \citep{LDstray}, and we believe it  occurs
at velocities that have the greatest fractional variation 
in $T_b$ across the FLS field: those of the
high-velocity cloud and those of the bright, low-velocity, narrow lines. 

In the bootstrapping technique that we adopted after much
experimentation, a GBT H\,I image was convolved with a 
circular Gaussian function to 36\arcmin\ angular resolution 
 and compared to  LD survey spectra at eight
locations near the field center.  Because stray radiation is expected to be
significant only at velocities $|V_{\rm LSR}| \lesssim
100$\,km\,s$^{-1}$, the observed GBT data were taken 
to be correct at  $|V_{\rm LSR}| > 125$\,km\,s$^{-1}$.  To compensate for the
unphysical estimate of the stray radiation at the velocity of the
narrow bright line, the stray spectrum was interpolated over the
velocity range $-9 \leq V_{\rm LSR} \leq +2.5$\,km\,s$^{-1}$
which contains this feature.  The stray column densities derived in this way 
vary from  $0.1 \times
10^{20}$\,cm$^{-2}$ for the data taken at LST 10$^{\rm h}$ to $1.2
\times 10^{20}$\,cm$^{-2}$ for the data taken at 23$^{\rm h}$.  On the
whole, the correction procedure appears to produce consistent results
for all six images when they are processed separately, so we averaged 
them to produce the final data
cube.  Figure 3 shows the uncorrected 21\,cm spectrum averaged over
the central part of all the maps of the FLS,
 and the average spectrum of the stray radiation which was subtracted from 
it, which has an equivalent $N_{\rm HI}= 4.9 \times 10^{19}$ cm$^{-2}$.

Work is now underway on projects that we hope will
eliminate the GBT's forward spillover lobe at 21\,cm, 
greatly improving its performance for H\,I observations, 
and making procedures like this unnecessary in the future.  We note that the
forward spillover lobe is important only at frequencies 1 -- 2 GHz.
Over most of its frequency range the GBT has an exceptionally clean 
antenna pattern.

\subsection{Error Estimates}

For $|V_{\rm LSR}|>125$ km s$^{-1}$ uncertainties in the final spectra
are dominated by random noise, while for velocities closer to zero the
uncertainties arise mainly from the correction for stray radiation.
An estimate of the latter was derived by comparing data cubes that had
quite different amounts of stray radiation removed.  The comparison suggests
that our correction for stray radiation may introduce an rms error of
0.1\,K for $|V_{LSR}|\gtrsim30$ km s$^{-1}$,  and as much as 
0.25\,K for the brighter emission near zero velocity.
At the eight positions across the final image where the stray
correction was derived, the integrated correction has a standard
deviation equivalent to $N_{\rm HI} = 
1.0 \times 10^{19}$\,cm$^{-2}$, which should
be a reasonable estimate for this error term.  We checked for temporal
variations in the stray correction by looking for systematic offsets
in corrected spectra as a function of time, but found nothing significant. 
The rms instrumental noise $\sigma \approx 0.08$\,K  was derived
from fluctuations in emission-free channels and is in good agreement
with theoretical estimates.  Table 2 summarizes these uncertainties.

\subsection{Creation of the $N_{\rm HI}$ Data Cubes}

After correction for stray radiation, the spectra were converted from
units of brightness temperature $T_{\rm b}(V)$ to H\,I column density
per channel $N_{\rm HI}(V)$.  This step requires knowledge of the
excitation temperature (called the spin temperature $T_{\rm s}$) of
the 21\,cm transition, which usually cannot be determined from the
21\,cm emission spectra themselves \citep{dl90, liszt01, fjl2004}.
Two conversions were therefore made: one assuming $T_{\rm s} =
10^4$\,K, which corresponds to optically thin emission, and the other
for $T_{\rm s} = 80$\,K, a value appropriate for diffuse clouds
\citep{savage77} and one which is more realistic for the narrow line
at low velocity which must arise from gas with $T_{\rm s} \leq 160 $ K
($\S3.1$).  Most of the H\,I emission has a $T_{\rm b}(V)$ of only a
few Kelvins, so the difference in assumed $T_{\rm s}$ affects the
total $N_{\rm HI}$ by only a few percent.  However,  at the velocities 
of the brightest lines, e.g., around $V_{\rm LSR} = -2$ km s$^{-1}$, 
$N_{\rm HI}(T_s=80)/N_{\rm HI}(thin)$ can be as large as 1.2. 
 In all likelihood, $T_{\rm s}$ varies across 
every 21cm H\,I spectrum \citep{liszt83}.  
Uncertainty in the conversion of $T_b$ to
$N_{\rm HI}$ affects the interpretation of 
dust-to-gas ratios in parts of the FLS field
(\S4).    The fundamental data cube of $T_{\rm
b}(\alpha, \delta, V)$ is available for anyone who wishes to make a
different conversion to $N_{\rm HI}(V)$.

The final data product is three cubes,
 available  via {\tt http://www.cv.nrao.edu/fls$\_$gbt} in FITS format. 
Data at the extreme ranges of velocity which contain no emission 
have been omitted from the final cubes.  The data before
correction for stray radiation are also available from F.~J.~Lockman
upon request.

\section{H\,I in the FLS Field}

\subsection{Selected H\,I Features}

H\,I spectra from the FLS field (e.g., Fig.~3) typically contain four 
components: (1) a broad line near zero velocity which 
contains most of the emission in most directions, (2) a bright, narrow 
line also near zero velocity which has a patchy though spatially 
correlated structure, (3) several clouds at intermediate negative 
velocities, and (4) a high-velocity cloud.

The high-velocity cloud appears 
over much of the FLS field, particularly at the
higher declinations.  It has a strong velocity gradient from 
 $V_{\rm LSR} = -190$\,km\,s$^{-1}$ in the east, where it is 
brightest, to  $V_{\rm LSR} = -155$\,km\,s$^{-1}$ in the west.  
This is part of high-velocity Complex C, a large
sheet of gas which extends $>100^\circ$ across the sky and is at least 5\,kpc
distant \citep{wvw91}.    The H\,I column density
integrated over the velocity of the cloud is shown in Figure 4.  The
peak $N_{\rm HI}$ of $6.9 \times 10^{19}$\,cm$^{-2}$ is about one-quarter 
of the total $N_{\rm HI}$ in that direction.

Several  H\,I clouds at an intermediate negative velocity are
visible in the FLS field (Figs.~5 and 6).  Properties 
 of these clouds are listed in Table 3.  Some are 
smaller than the GBT beam. They 
contribute 10\% -- 25\% to the total $N_{\rm HI}$ in their directions.
Unlike high-velocity clouds, which do not show FIR emission \citep{wb86}, 
intermediate-velocity clouds are highly
correlated with $I_{100\mu{\rm m}}$, though sometimes 
with a smaller $I_{100\mu{\rm m}}/N_{\rm HI}$ ratio 
than quiescent gas \citep{deulburton}.  Of
special interest is the rather simple cloud shown in Figure 6, which
seems to be  part of the Draco nebula, an object $\ga 600$\,pc distant. 
The Draco nebula is quite distinct at $100\mu$m and is seen in absorption 
against the soft X-ray background \citep{burrows, herbstm, gladders, penprase}.

Finally, the integrated $N_{\rm HI}$ image of Fig.~7 shows an arc of
emission to the south-west (lower right) which arises in a narrow line
at $V_{\rm LSR} = -2$\,km\,s$^{-1}$.  The line-width of 
2.7\,km\,s$^{-1}$ FWHM implies that the gas has a kinetic temperature
$T\leq160$\,K.  This is the brightest line in the field, reaching a peak
 $T_{\rm b} = 26$~K, corresponding to $\tau_{HI} = 0.4$ and 
$N_{\rm HI} = 1.8 \times 10^{20}$\,cm$^{-2}$ for $T_s = 80$ K. 
Most of the emission in the arc appears to be resolved 
 by the beam of the GBT, though there may be small unresolved components.

\subsection{Integral Properties}

The H\,I column density for $T_{\rm s} = 80$\,K integrated over all
velocities is shown in Figure 7.  The visible structure is caused 
by specific interstellar objects, many of which have been discussed above.
   Table 4 lists values of  $N_{\rm HI}$ averaged over 
 various areas centered on the FLS central position.   
Over 9\,deg$^2$ the total $N_{\rm HI}$ varies by  
only a factor of two, with an rms scatter  13\% of the mean.  
As  in most directions at high Galactic latitude, 
structure in H\,I over the FLS field is highly spatially correlated; 
significant fluctuations in total $N_{\rm HI}$ do not  arise randomly 
or from  tiny clouds (e.g., \citet{schlegel, miville, fjl2004}).

\section{Infrared -- H\,I Correlations}

\citet{schlegel} hereafter SFD, have derived the 
dust column density across the sky from temperature-corrected $100\mu$m 
data, and have expressed it as a reddening $E(B-V)$.  
 Figure 8 shows their values of $E(B-V)$ over the FLS
field at full  resolution.  
(The SFD maps have an angular resolution of $6\farcm1$, but 
 quantitative comparisons with H\,I are made 
using the SFD maps smoothed to the angular resolution of the H\,I maps.) 
Most dust  features have exact
counterparts in H\,I.  The correlation is shown explicitly in the 
left panel of Figure 9, where the FIR-derived reddening $E(B-V)$ 
is plotted against  $N_{\rm HI}$. 

 The H\,I spectra consist of many components, and 
correlations performed between $E(B-V)$ and 
$N_{\rm HI}$ in various velocity intervals  show that high-velocity Complex C
must have a $100\mu$m emissivity per H\,I atom at least an order of 
magnitude smaller than the other H\,I components.  
For Complex C, the $3\sigma$ limit is 
$I_{100\mu\rm{m}}/N_{\rm HI} < 5 \times 10^{-22}$ MJy sr$^{-1}$ cm$^2$,  
while the ratio ranges 
between 50 and 70 in the same units 
for H\,I at other velocities in the FLS field, values comparable to those 
 found in other studies (e.g., \citet{heiles99, lagache00}). 
This confirms previous findings of only upper limits on
$I_{100\mu{\rm m}}$ from high-velocity clouds \citep{wb86, wvw97}.
Complex C has low metallicity and contains little dust \citep{tripp03}
which can account for its low FIR emissivity.
The right panel of Figure 9 shows the $E(B-V)$ vs.~$N_{\rm HI}$ when 
 H\,I associated with the high-velocity cloud is not included.  
The scatter is greatly reduced.

The lines in Fig.~9 are fit to $E(B-V) = a_0 + a_1N_{\rm
HI}(T_s=80$ K) and the coefficients $a_0$ and $a_1$ are listed in Table 5. 
  The lower panel of Figure 8 shows $N_{\rm HI}$ 
for $T_{\rm s} = 80$ K without the high-velocity cloud.  
In contrast to the H\,I map of Fig.~7, which includes all H\,I, 
 this $N_{\rm HI}$ map appears nearly identical to the dust map, 
though there remain interesting anomalous regions.    A map of the ratio 
 $E(B-V) / N_{\rm HI}$ over the FLS field is shown in Figure
10.  The ratio varies between 0.9 and 1.5 with a mean of $1.17\pm0.09 
\times 10^{-22}$ mag cm$^{2}$. 

\subsection{Infrared -- H\,I Variations}

A detailed analysis of the gas and dust in the FLS field is beyond the scope 
of this paper,  for there issues of dust properties and temperature, the 
conversion between $I_{100\mu\rm{m}}$ and $E(B-V)$, 
and the dust associated with ionized gas 
\citep{lagache99, lagache00}. Here we will consider only what 
can be learned from the  H\,I about possible sources of the 
 variation in $E(B-V)/N_{\rm HI}$ across the FLS field, for 
some variations in the ratio seem tied to the properties of the 21cm line. 

A significant aspect of Fig.~10 is that spatial 
structure in the dust-to-gas ratio is highly correlated, a fact that 
has been known for some time (e.g., SFD).
Perhaps surprisingly, high dust-to-gas ratios come 
preferentially from areas of high $N_{\rm HI}$.  One can see 
in the right panel of Fig.~9 that high $N_{\rm HI}$ directions 
are redder than a strict linear relationship would suggest.  Table 5 
includes results of a linear fit to only that data with 
$E(B-V)<0.03$, equivalent to  $N_{\rm HI} \lesssim 2.5 \times 10^{20}$ 
cm$^{-2}$; the slope is significantly flatter than when all 
reddenings are included.  But higher reddening 
per unit $N_{\rm HI}$ does not simply track areas of 
high $N_{\rm HI}$:  the greatest 
$N_{\rm HI}$ in the FLS field is in the upper left, yet 
it has an average value of $E(B-V) / N_{\rm HI}$ and is not 
at all conspicuous in Fig.~10.  We find, instead, 
that the best predictor of a high $E(B-V) / N_{\rm HI}$ over the 
FLS field is the peak brightness temperature in the 21cm line, as shown 
in Figure 11.  There seems to be excess reddening when 
 spectra have a peak $T_{\rm b} \ga 12$ K, and the discrepancy increases 
as the line gets brighter.  The spectra with the highest $T_b$ are 
those toward the `arc' at $V_{\rm LSR} \approx 0$.  Similar, 
and possibly related line components cover much of the map. 
Assuming that the values of $E(B-V)$ are accurate, 
we consider three possibilities which might explain the increased reddening 
in this component.

First, the values of  $N_{\rm HI}$ we derive would be 
underestimates if the gas in the brighter lines is 
 cooler than the assumed $T_s=80 $. 
For example, if the brightest location in the 
`arc' is evaluated for $T_s=38$ K instead of 80 K, 
its derived  $N_{\rm HI}$ would be increased enough to give 
it an average dust-to-gas ratio.  This however, does not help 
other directions on the map very much: for $T_s = 38$ K 
there still remain large correlated areas of high dust-to-gas ratio, 
and the standard deviation of the ratio across the map is reduced 
by only $15\%$ from the $T_s=80$ K values.  Adopting even lower values of 
$T_s$ would create areas of below-average ratios in the map.  We believe
that the features of Figs.~10 and 11 cannot simply be the result of 
an incorrect $T_s$. 

Second, it is possible that  $N_{\rm HI}$ does not reflect the total 
 $N_{\rm H}$ because some gas is in the form of  H$_2$.  
While it has long been known  that significant amounts of molecular
hydrogen are seen in directions with total $N_{\rm HI} \geq 5 \times
10^{20}$\,cm$^{-2}$ \citep{bsd78}, the total $N_{\rm HI}$ 
along most sight-lines is most likely  an integration 
over gas of very different properties, diffuse and dense 
 (e.g., \citet{rachford}), and  H$_2$ might be found 
in individual clouds at smaller values of  $N_{\rm HI}$.  
If the  `arc' in the FLS field has 
a line-of-sight size similar to its tranverse size, then it has 
an average H\,I density $\langle n \rangle \approx 100\  d_{100}^{-1} $
cm$^{-3}$, where $d_{100}$ is its distance in hundreds of pc.  
Although the distance to this gas is not known, it is almost certainly 
local, within a few hundred pc, and thus 
$\langle n \rangle > 30$ cm$^{-3}$.  At these densities, 
and for its $N_{\rm HI}=1.8 \times 10^{20}$ cm$^{-2}$, 
theoretical calculations show that it should 
have $N(\rm H_2)/N(\rm H\,I) >  10\% $ \citep{lisztlucas2000, liszt2002}.
This amount would be sufficient to account for the apparent excess 
of dust in the feature and in similar objects in the FLS field.  
In this interpretation, the brightest H\,I lines are coming from the 
densest clouds, which have some molecules, an inference supported
by other IR--HI correlations \citep{hrk88}.

Finally, it is possible that the dust-to-gas ratio, or the 
character of the dust, or its infrared emission, really does 
vary as illustrated in Fig.~10, and that the connection to the 
H\,I spectrum peak $T_{b}$ simply identifies a particular interstellar 
cloud, and does not have physical significance.  In this regard 
it is interesting that the mean dust-to-gas ratio in the FLS field 
derived from the SFD reddening values 
implies that one magnitude of reddening requires an  $N_{\rm HI}$ 
of $8.6 \times 10^{21}$ cm$^{-2}$, almost $50\%$ higher than the 
canonical value of $5.8 \times 10^{21}$ cm$^{-2}$ \citep{bsd78}, 
which resulted from direct observations in the ultraviolet and 
optical.  In fact, if we adopt the \citet{bsd78} relationship as the 
correct one, then the anomaly in the FLS field is that there is too 
little dust except in the arc and related regions.

\section{Final Comments}

The H\,I data presented here should be useful for estimating the
effects of Galactic foreground emission on the Spitzer FLS data.  We
find that for $N_{\rm HI} \la 2.5 \times 10^{20}$ cm$^{-2}$ there is a
linear relationship between $E(B-V)$ derived from FIR emission and
$N_{\rm HI}$, provided that high-velocity H\,I is omitted.  Directions
where the 21\,cm line has a peak brightness $T_{\rm b} \ga 12$ K show
above-average reddening per unit $N_{\rm HI}$ and may 
contain some H$_2$.  The correlation between excess dust per 
H\,I atom and peak $T_{\rm b}(V)$, rather than with the total  $N_{\rm HI}$,
may arise because a high $T_{\rm b}(V)$ can signify   pileup of gas
at a specific volume in space, and hence a high density and local 
shielding, whereas a high total $N_{\rm HI}$ 
can result from the sum of many unrelated, low density  regions. 
Comparison of the individual H\,I spectral
components with infrared data from the Spitzer FLS may find 
variations in the FIR emissivity of different spectral features, and possibly
even reveal a component connected with the high-velocity cloud.

\acknowledgments
We thank D.P. Finkbeiner, J.R. Fisher, P. Martin, 
R. Norrod, and S. Srikanth for helpful discussions.


\begin{figure}
\includegraphics[angle=-90,scale=0.6]{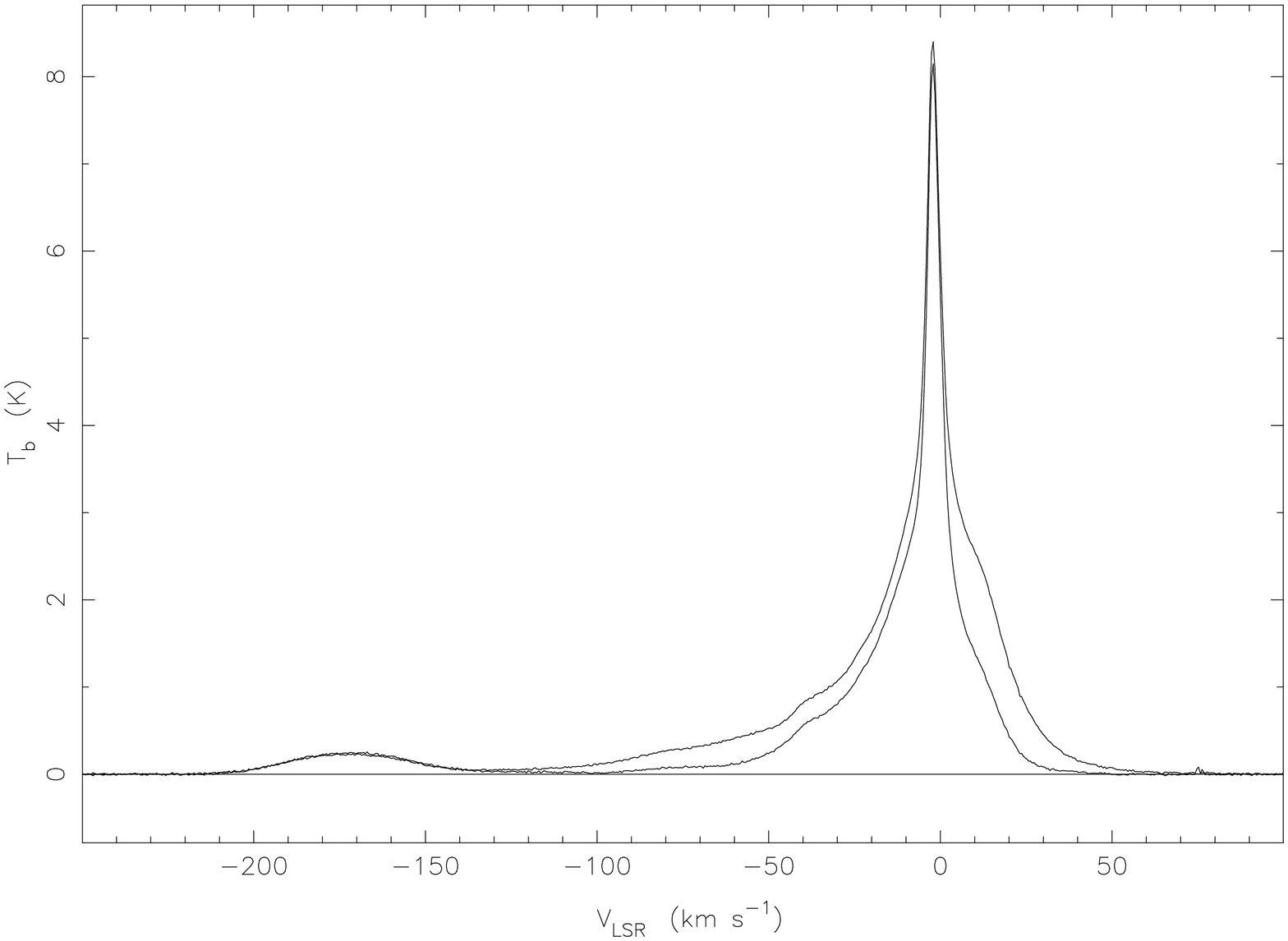}
\caption{These H\,I spectra averaged over our $3\arcdeg \times
3\arcdeg$ FLS area were obtained at two different local sidereal
times.  The spectrum with the stronger emission was obtained at $LST =
23^{\rm h}$ while the weaker spectrum was obtained near $LST = 10^{\rm
h}$.  Their differences are attributable entirely to the changing
amount of stray radiation coming through the forward spillover lobe of
the GBT.  Stray radiation is not expected to be significant at
velocities $|V_{\rm LSR}| > 125 $\,km\,s$^{-1}$;   note that the
high-velocity emission near $ V_{\rm LSR} = -175$\,km\,s$^{-1}$ 
does not vary.}
\end{figure}
\clearpage

\clearpage
\begin{figure}
\epsscale{0.5}
\includegraphics[angle=-90,scale=0.65]{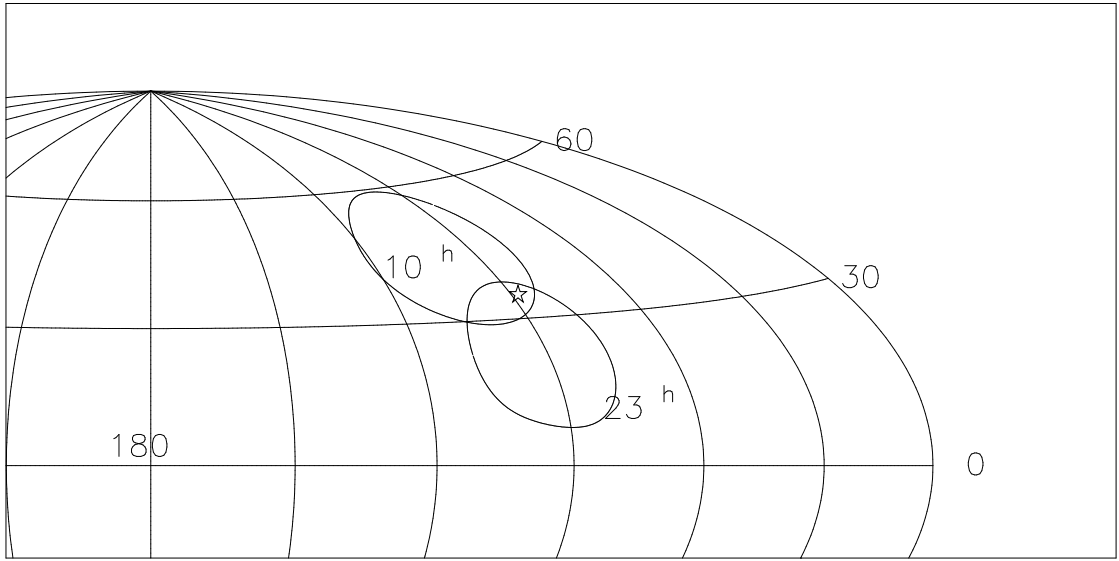}
\caption{The projections of the rim of the 
8\,m diameter GBT subreflector onto Galactic 
coordinates at 10$^{\rm h}$ LST and 23$^{\rm h}$ LST when the main beam is 
pointed in the direction of the FLS survey region (marked with a
star).  Emission from H\,I which enters the receiver from just outside the 
subreflector rim is expected to have a strong diurnal variation as the 
subreflector moves with respect to the Galactic plane, an effect seen 
in the spectra of Fig.~1.
}
\end{figure}
\clearpage

\clearpage
\begin{figure}
\includegraphics[angle=-90,scale=0.6]{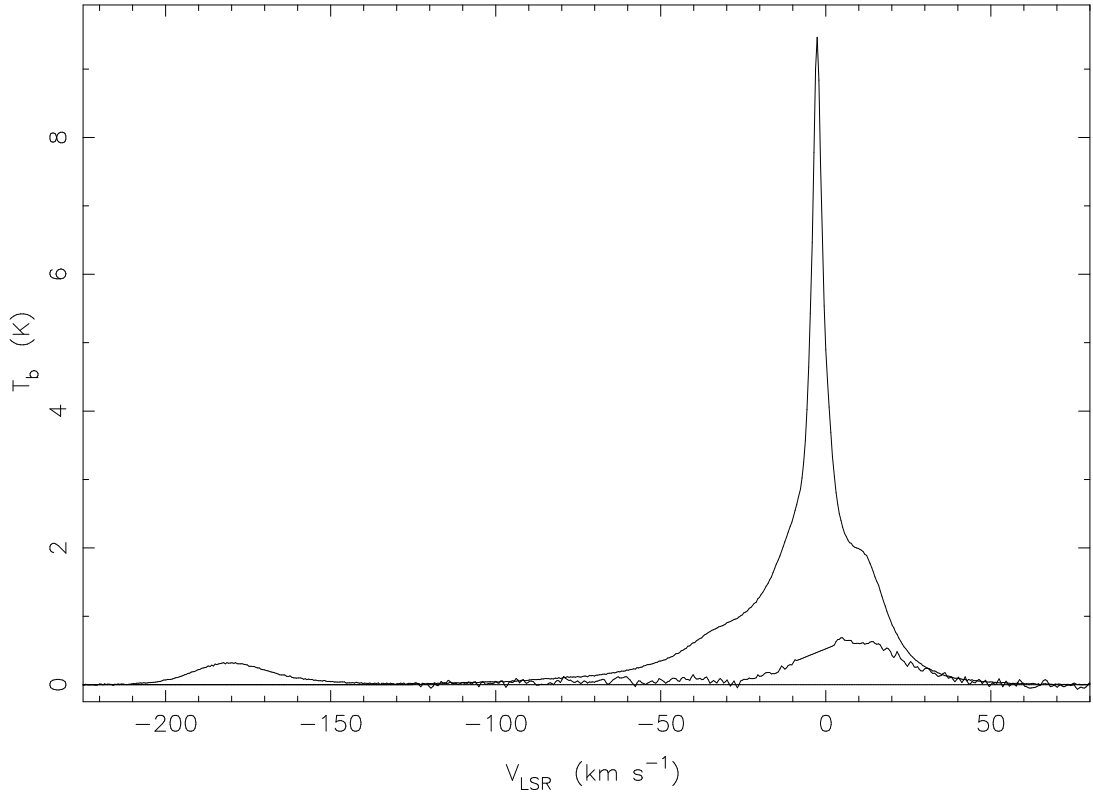} 
\caption{The average of GBT spectra from all maps over the inner 
two square degrees of the FLS field, and
the portion of the emission that we attributed to stray radiation.  Note the
linear interpolation of the stray spectrum between $V_{\rm LSR} =
-9$\,km\,s$^{-1}$ and $ V_{\rm LSR} = +2.5$\,km\,s$^{-1}$.  
The stray radiation contains an equivalent $N_{\rm HI}=4.9 \times 10^{19}$
 cm$^{-2}$, which amounts to  $17\%$ of the total observed signal.
}
\end{figure}
\clearpage

\clearpage
\begin{figure}
\epsscale{0.5}
\includegraphics[angle=0,scale=1.]{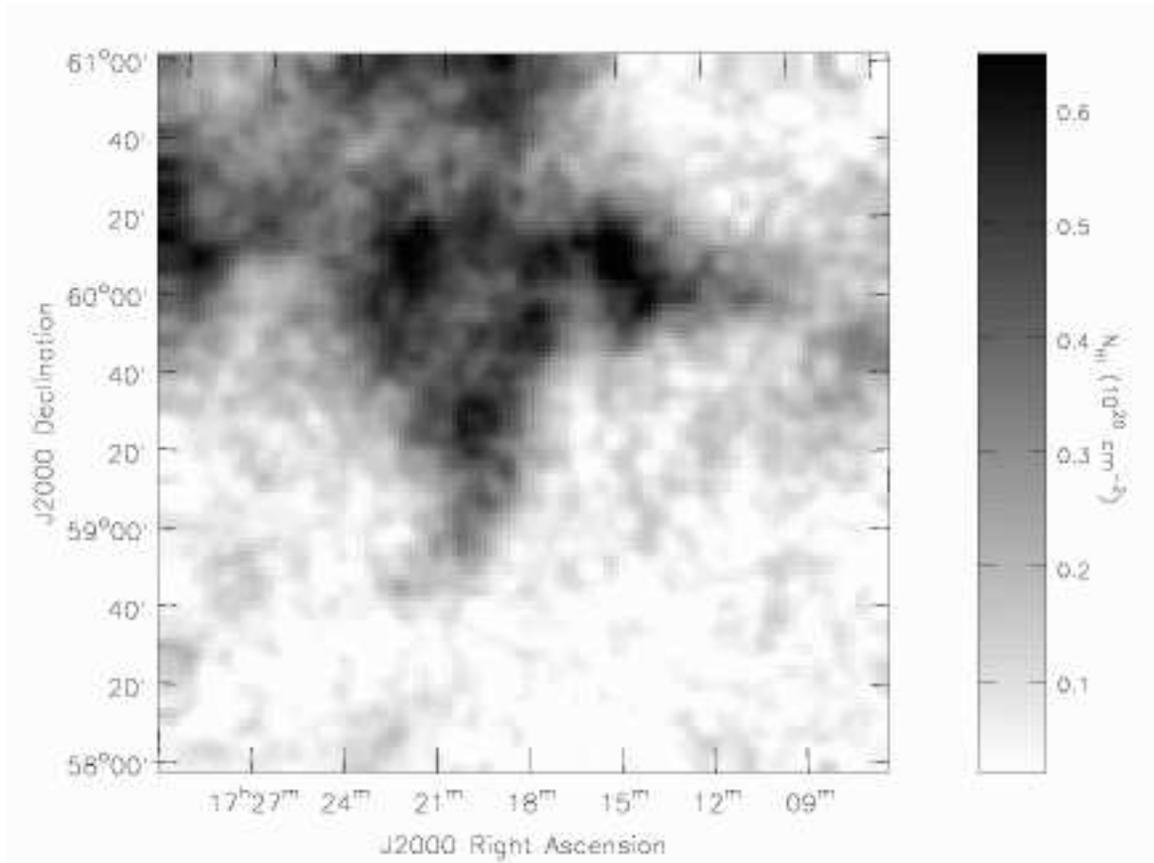}
\caption{H\,I in the FLS field integrated over $-230 \leq V_{\rm LSR} 
\leq -130$\,km\,s$^{-1}$ showing emission
arising from  high-velocity cloud Complex C.
}
\end{figure}
\clearpage

\clearpage
\begin{figure}
\includegraphics[angle=0,scale=1.]{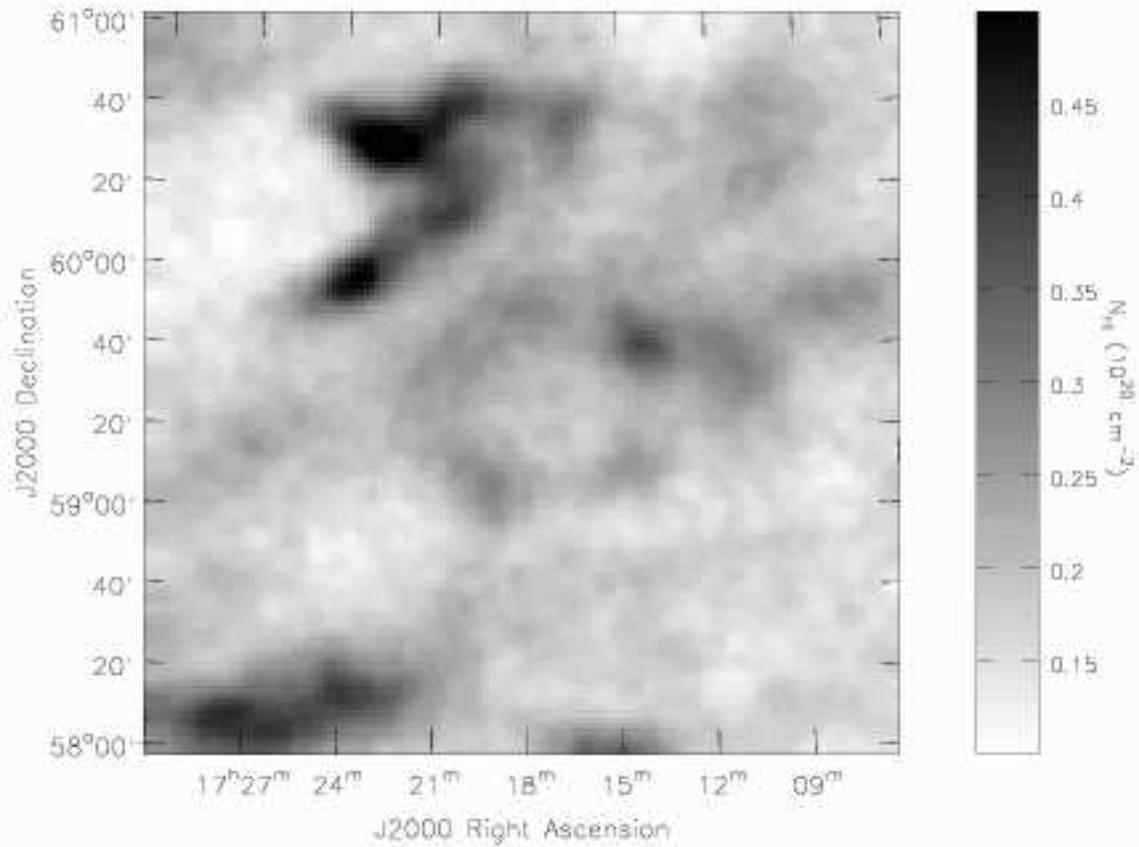}
\caption{H\,I in the FLS field integrated over $-45 \leq
V_{\rm LSR} \leq -35$\,km\,s$^{-1}$ showing 
several of the intermediate-velocity clouds.  }
\end{figure}
\clearpage

\clearpage
\begin{figure}
\epsscale{0.5}
\includegraphics[angle=-0,scale=1.]{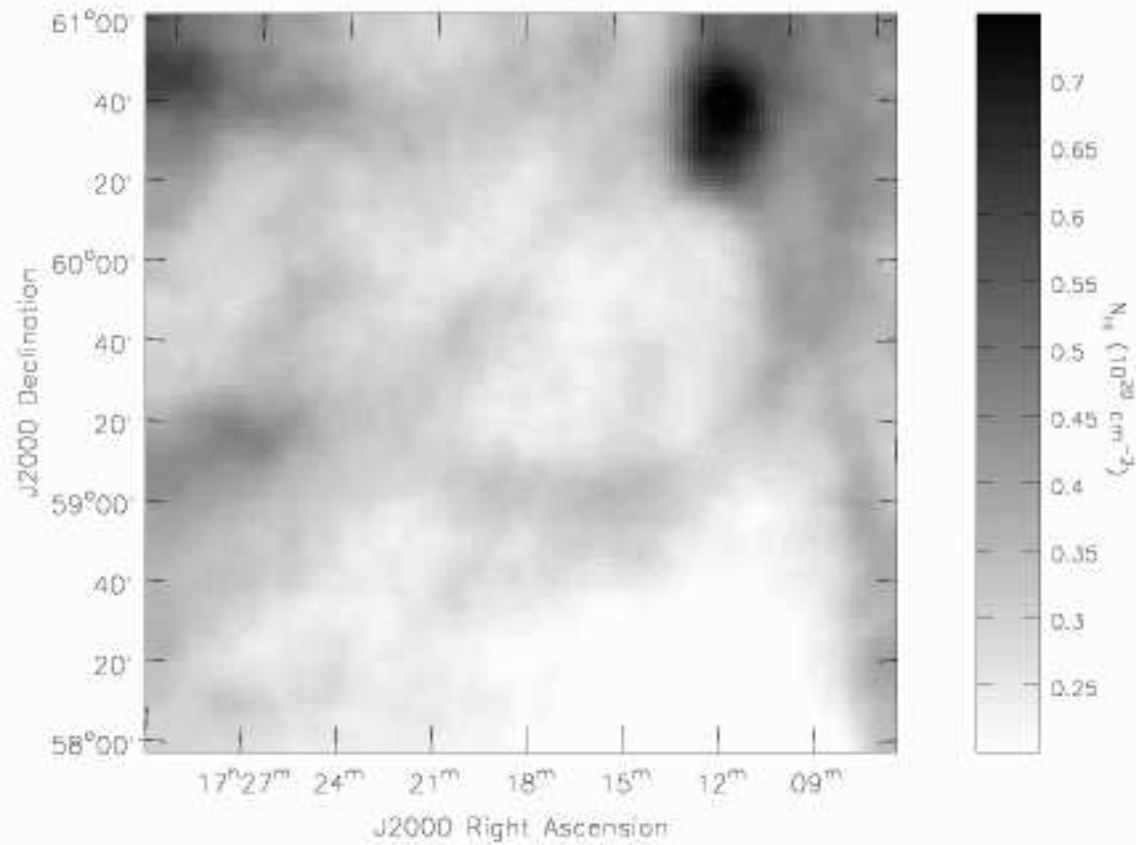}
\caption{H\,I in the FLS field integrated over $-29 \leq
V_{\rm LSR} \leq -17$\,km\,s$^{-1}$ highlighting emission from the small
intermediate-velocity cloud at $\delta = 60\arcdeg 41\arcmin$ 
 associated with the Draco nebula.  }
\end{figure}
\clearpage

\clearpage
\begin{figure}
\epsscale{0.8}
\includegraphics[angle=0,scale=1.25]{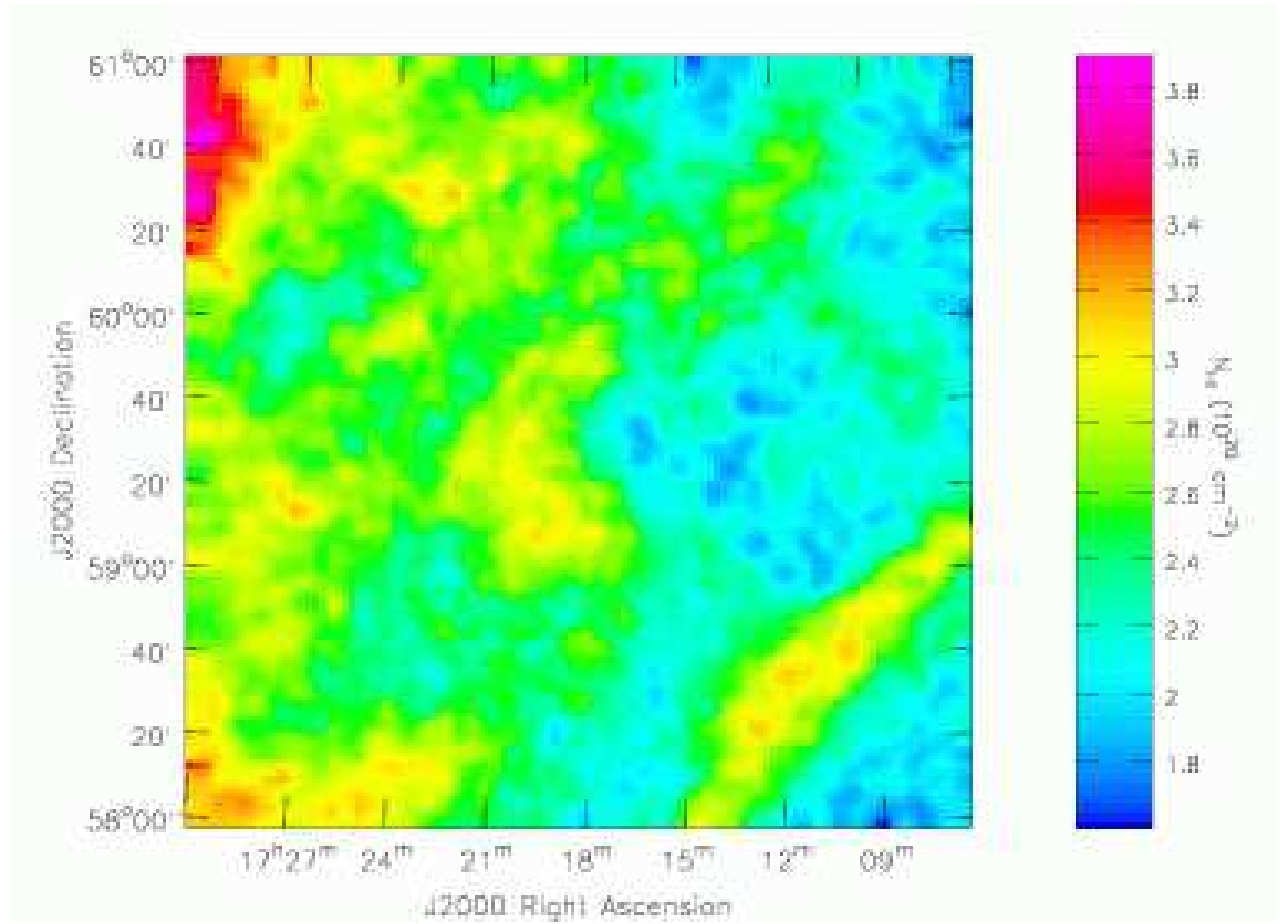}
\caption{The $N_{\rm HI}$ over the FLS field 
integrated over all velocities after correction 
for stray radiation, evaluated for an assumed excitation temperature 
$T_{\rm s} =  80$\,K.  The `arc' discussed in the text is the 
feature to the lower right.  It arises from a narrow, bright, 
low-velocity line.
}
\end{figure}
\clearpage

\clearpage
\begin{figure}
\epsscale{0.5}
\includegraphics[angle=0,scale=0.75]{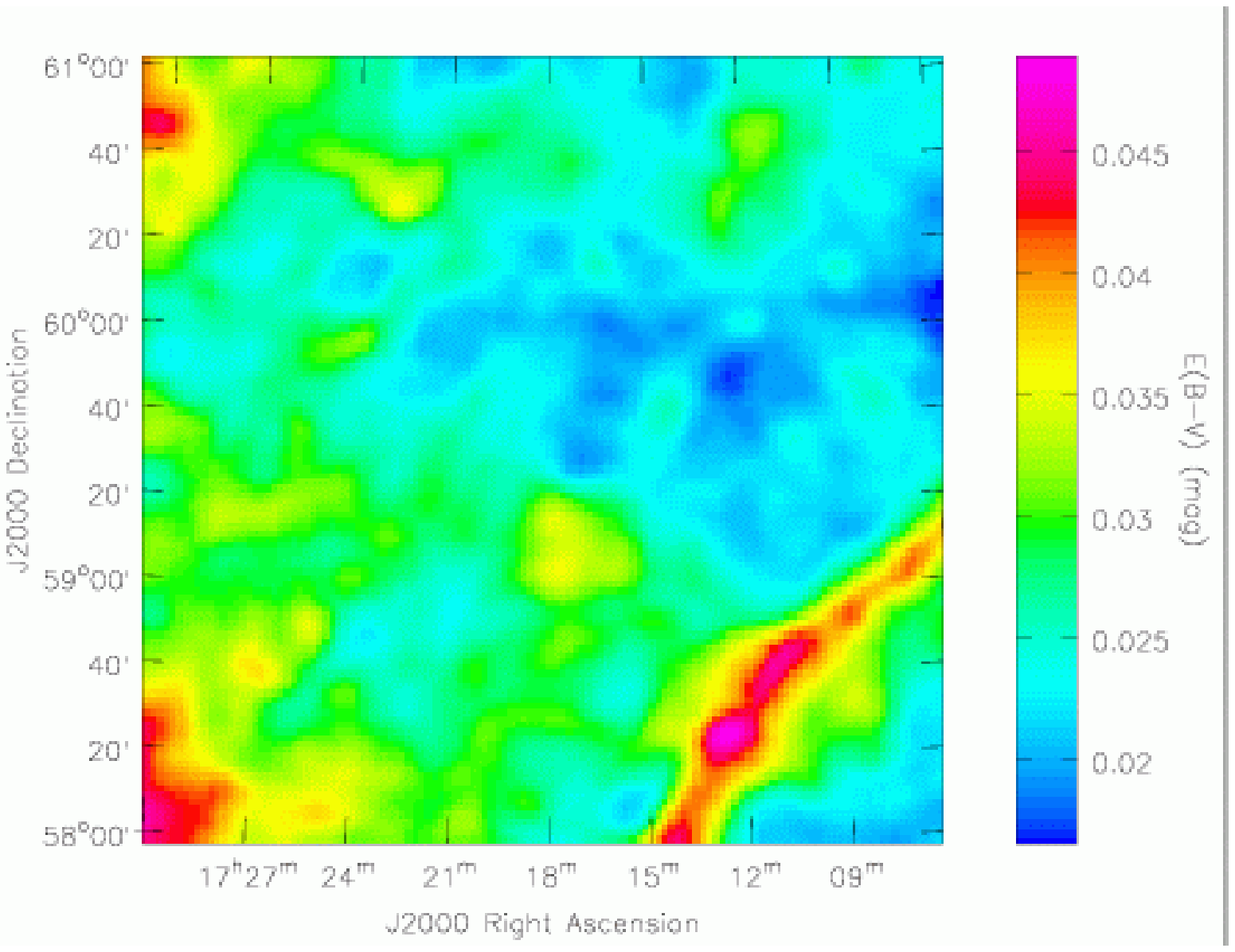}
\includegraphics[angle=0,scale=0.75,origin=c]{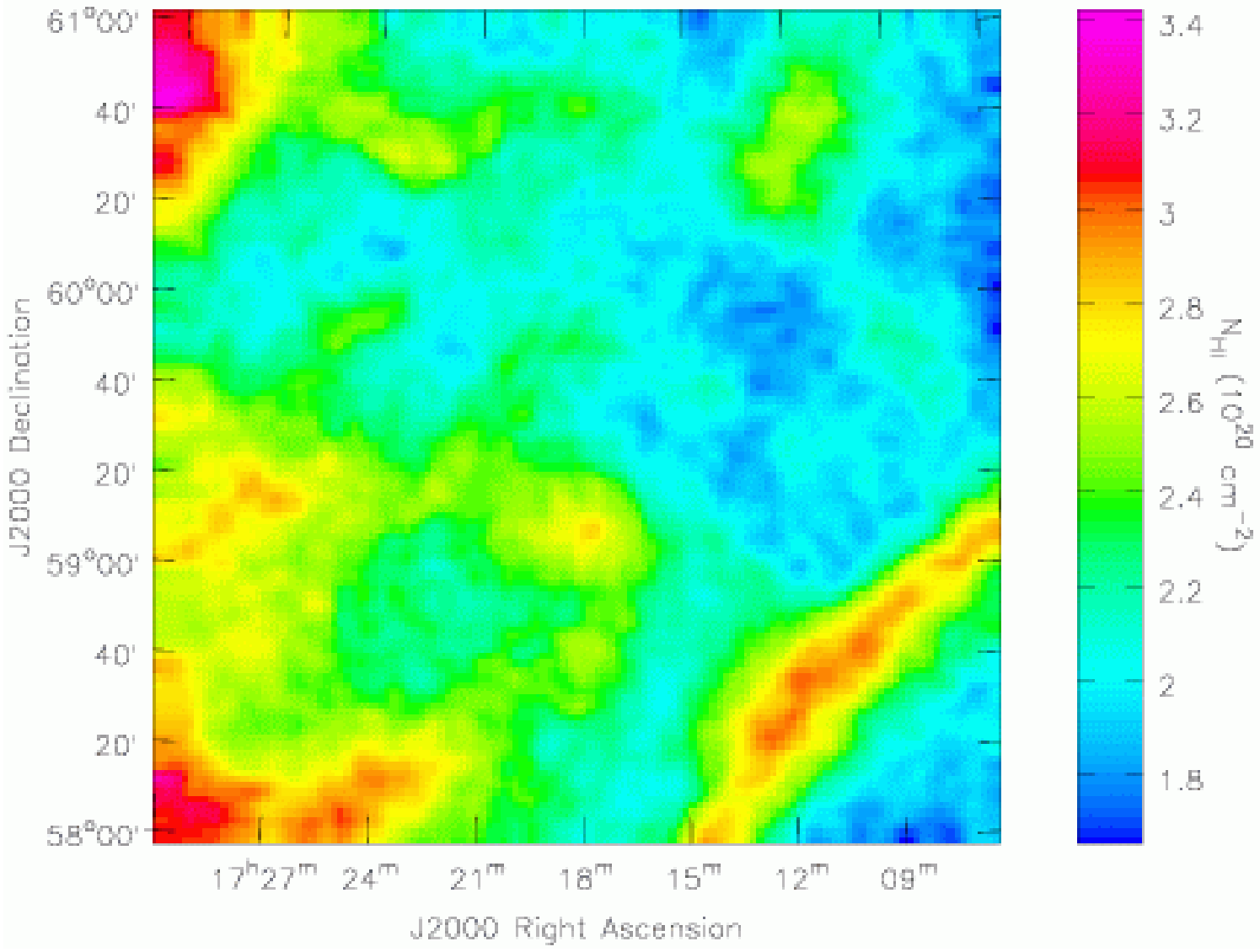}
\caption{Top Panel:  $E(B-V)$  over the FLS field 
derived by  \citet{schlegel} from temperature-corrected 
${100\mu{\rm m}}$ data at $6\farcm1$ angular 
resolution.  Bottom Panel: 
Column density $N_{\rm HI}$ for $T_{\rm s} = 80$ K 
from the GBT observations integrated over $V_{\rm LSR} > -100$  km s$^{-1}$
which excludes the high-velocity cloud. 
The bottom panel should be compared to Figure 7, which includes 
H\,I from the high-velocity cloud, to illustrate the absence of 
FIR emission from high-velocity clouds.
}
\end{figure}
\clearpage

\clearpage
\begin{figure}
\epsscale{0.5}
\includegraphics[angle=-90,scale=0.6]{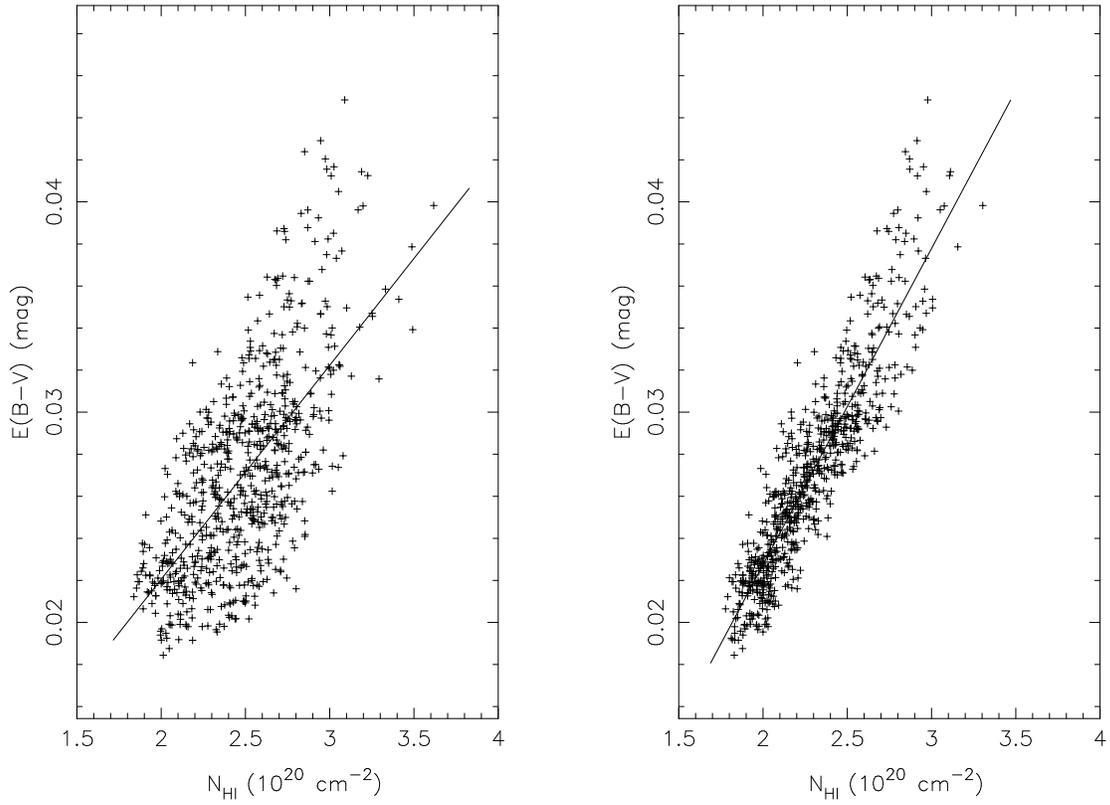}
\caption{The amount of dust expressed as a reddening $E(B-V) $
from  \citet{schlegel} vs.  GBT H\,I column density for 
the FLS field, where the H\,I column is derived for $T_{\rm s} = 80$ K.  
The dust data have been convolved to the angular resolution 
of the H\,I, and to reduce clutter only every second data 
point is plotted.  The left panel shows the correlation for H\,I at 
all velocities, while the right panel shows the correlation omitting 
H\,I from the high-velocity cloud.  Solid lines are a linear fit 
to $E(B-V)=a_0+a_1N_{\rm HI}$; coefficients are given in Table 5. 
}
\end{figure}
\clearpage

\clearpage
\begin{figure}
\epsscale{0.5}
\includegraphics[angle=-0,scale=.8]{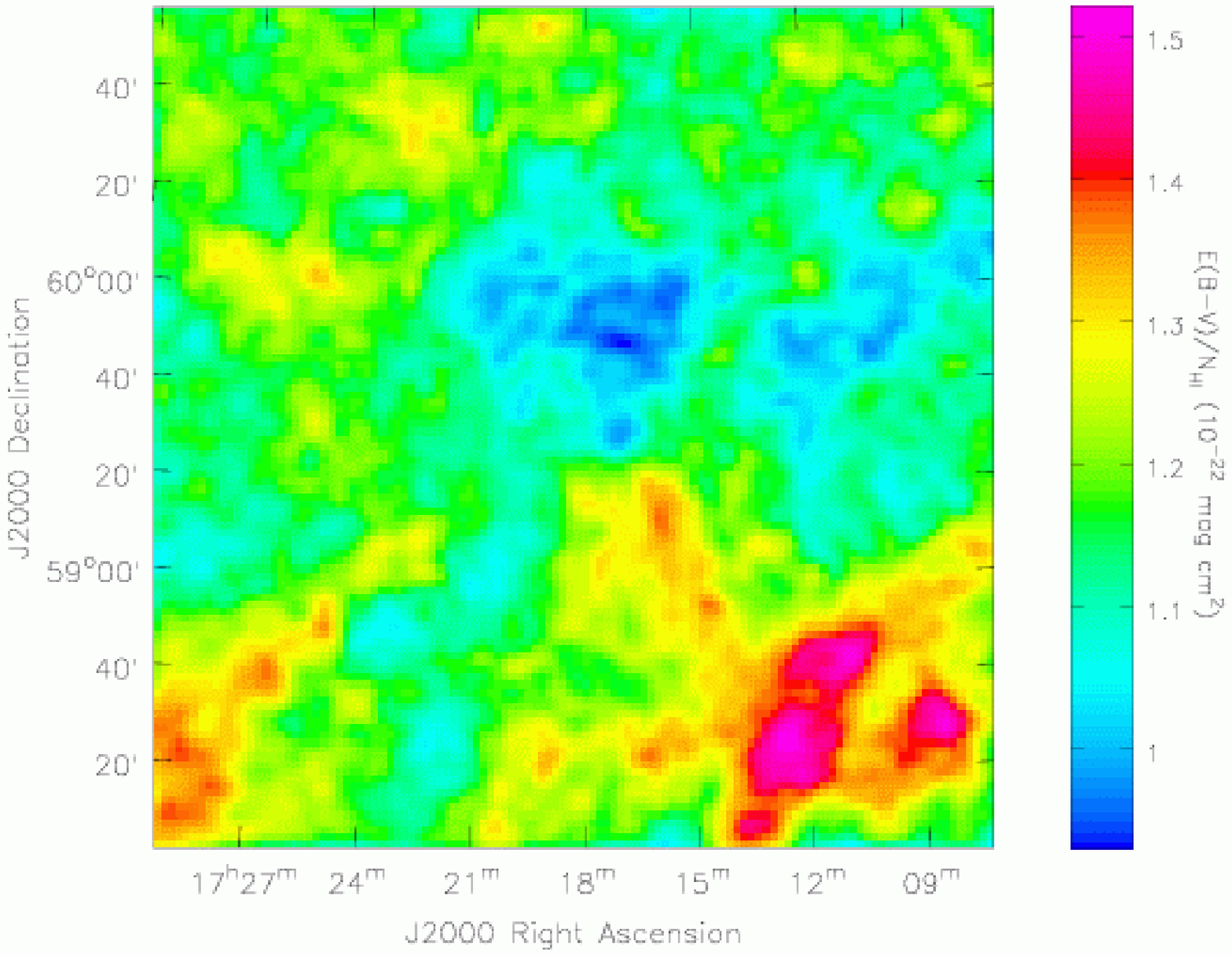}
\caption{A map of the ratio  of reddening to H\,I, 
$E(B-V)/N_{\rm HI}$, over the FLS field.  The reddening
 is from  \citet{schlegel} 
smoothed to the angular resolution of the H\,I data, 
and the H\,I is for $T_s =80$ K and omits emission associated 
with  the high-velocity cloud.  
Areas with the highest ratios are generally not those with the largest 
$N_{\rm HI}$ but those with the largest $T_{\rm b}(V)$:  
the largest $N_{\rm HI}$ occurs in the upper 
left of the map (see Fig.~8), a region of average $E(B-V)/N_{\rm HI}$, 
while the most dust per unit $N_{\rm HI}$ is in the `arc' feature 
and related emission at the lower right. 
}
\end{figure}
\clearpage

\clearpage
\begin{figure}
\includegraphics[angle=-90,scale=0.65]{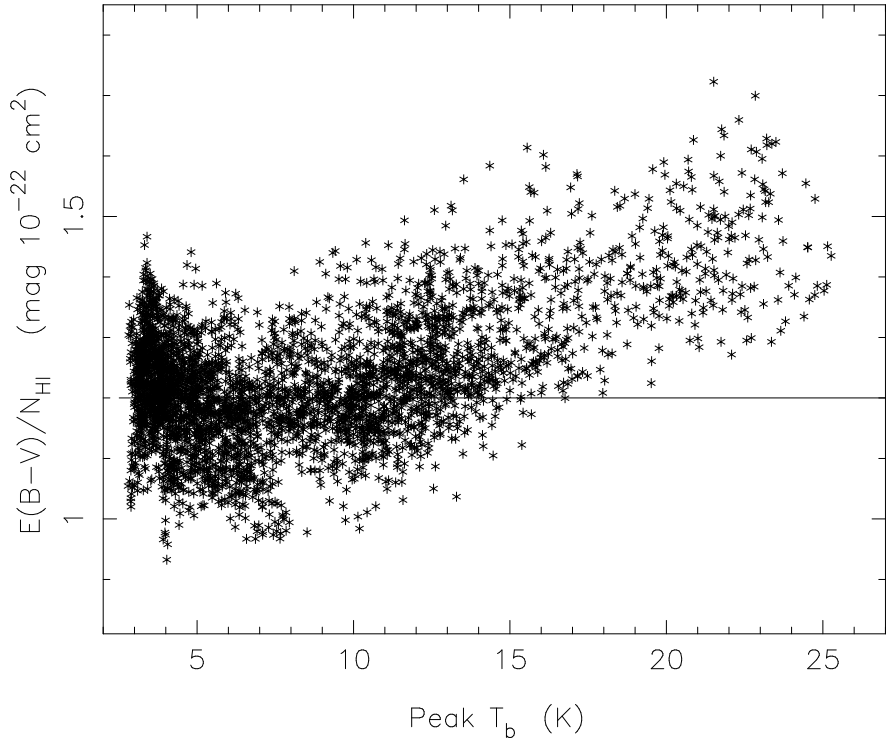}
\caption{The ratio $E(B-V)/N_{\rm HI}$ for $T_s=80$ K,  
vs. the peak brightness temperature in the H\,I line.  
To reduce clutter only every second point is plotted. 
When the 21cm spectrum has a component with $T_b \ga 12$ K, the reddening 
per H\,I atom is higher than the average value, 
$1.2 \times 10^{-22}$ mag cm$^2$, found 
for the most transparent lines of sight.  This suggests that for 
$T_{\rm b} > 12$ K the derived value of $N_{\rm HI}$ is not representative of 
the true $N_{\rm H}$  because H\,I is cooler than 
the assumed $T_s=80$ K,  or  because that direction contains some $H_2$.  
}
\end{figure}
\clearpage
 
\begin{deluxetable}{lc}
\tablecolumns{2}
\tablenum{1}
\tablewidth{0pt}
\tablecaption{GBT  H\,I Map of the FLS Field}
\tablehead{ }
\startdata 
 Field Center (J2000) &  $17^{\rm h}18^{\rm m} +59\arcdeg30\arcmin$ \\
 Field Center (Galactic) & 88\fdg32 +34\fdg89 \\
 Map Size (J2000) & $3\arcdeg \times 3\arcdeg$ \\
 Effective Angular resolution & $9\farcm8$ \\ 
 Pixel Spacing & $1\farcm5$ \\
 $V_{\rm LSR}$  Coverage (km s$^{-1}$)  & $-314$ to $+213$\\
 Velocity Resolution &  0.62 km s$^{-1}$ \\ 
 Channel Spacing & 0.52 km s$^{-1}$ \\
 
\enddata
\end{deluxetable}

\begin{deluxetable}{ll}
\tablecolumns{2}
\tablenum{2}
\tablewidth{0pt}
\tablecaption{Uncertainties in the Final Data}
\tablehead{
\colhead{} & \colhead{}   \\ 
}
\startdata 
Noise  & 0.08 K ($1\sigma$) \\
Stray Radiation  & 0.1 -- 0.25 K ($1\sigma$) \\
Total $N_{\rm HI}$ error & $1.1 \times 10^{19}$ cm$^{-2}$ ($1\sigma$) \\
 
\enddata

\end{deluxetable}

\vfill \eject

\begin{deluxetable}{lccrr}
\tablecolumns{5}
\tablenum{3}
\tablewidth{0pt}
\tablecaption{H\,I Features in the FLS field}
\tablehead{
\colhead{Object} & 
\colhead{$\alpha,\delta({\rm J2000})$} & \colhead{Peak $N_{\rm HI}$ }  
& \colhead{$V_{\rm LSR}$}  & \colhead{FWHM }\\
\colhead{} & \colhead{} & \colhead{(cm$^{-2}$)} & \colhead{(km s$^{-1})$} & 
  \colhead{(km s$^{-1}$)} \\
\colhead{(1)} & \colhead{(2)} & \colhead{(3)} & \colhead{(4)}  
& \colhead{(5)}\\

} 
\startdata 
Complex C  & $17^{\rm h}15^{\rm m}00^{\rm s} +60^\circ09'$ &
 $6.9 \times 10^{19}$ & $-190$ ~~~~~~~ & 20.0 ~~~~~~~ \\
IVC 1 &  $17^{\rm h}22^{\rm m}28^{\rm s} +60^\circ30'$ & 
 $5.6 \times 10^{19}$ & $-41$ ~~~~~~~ & 5.1 ~~~~~~~ \\
IVC 2 &  $17^{\rm h}23^{\rm m}24^{\rm s} +59^\circ58'$ &
 $5.0 \times 10^{19}$ & $-41$ ~~~~~~~ & 5.3 ~~~~~~~ \\
IVC 3 &  $17^{\rm h}27^{\rm m}39^{\rm s} +58^\circ06'$ &
 $4.4 \times 10^{19}$ & $-34$ ~~~~~~~ & 15.8 ~~~~~~~ \\
IVC 4 (Draco) &  $17^{\rm h}11^{\rm m}16^{\rm s} +60^\circ40'$ &
 $7.8 \times 10^{19}$ & $-23$ ~~~~~~~ & 5.4 ~~~~~~~ \\
Arc &  $17^{\rm h}10^{\rm m}30^{\rm s} +58^\circ41'$ &
 $1.8 \times 10^{20}$ & $-2$ ~~~~~~~ & 2.7 ~~~~~~~ \\
 
\enddata 
\tablecomments{The quantities refer to the direction of greatest 
$N_{\rm HI}$.}
\end{deluxetable}

\vfill \eject

\begin{deluxetable}{ccccc}
\tablecolumns{5}
\tablenum{4}
\tablewidth{0pt}
\tablecaption{Statistics on total $N_{\rm HI}$ ($T_{\rm s} = 80$ K) 
across the FLS Field}
\tablehead{
\colhead{Area\tablenotemark{a}} & \multicolumn{4}{c} {
   $ N_{\rm HI}\  (10^{20}$ cm$^{-2})$ } \\
\colhead{(deg$^2$)} &  \colhead{min} & \colhead{max} &
\colhead{avg} & \colhead{Std Dev} \\
\colhead{(1)} &\colhead{(2)} & \colhead{(3)} & 
\colhead{(4)} & \colhead{(5)} \\
}

\startdata 
0.02\tablenotemark{b} & & & 2.36 & \\
1 & 1.8 & 3.1 & 2.51 & 0.28 \\
4 & 1.8 & 3.2 & 2.45 & 0.25 \\
9 & 1.7 & 3.9 & 2.48 & 0.32 \\
\enddata
\tablenotetext{a}{Square region centered on the FLS.}
\tablenotetext{b}{Single pointing at J2000 $\alpha = 17^{\rm h}
  18^{\rm m},\,\delta = +59^\circ 30'$. 
}
\end{deluxetable}
\ 
\vfill \eject

\begin{deluxetable}{ccccc}
\tablecolumns{5}
\tablenum{5}
\tablewidth{0pt}
\tablecaption{$E(B-V) = a_0 + a_1N_{\rm HI}$~ (for T$_s = 80$ K)}
\tablehead{
\colhead{H\,I Velocity Range } & $E(B-V)$ & \colhead{a$_0$}  & 
\colhead{ a$_1$}  & \colhead{Std. Dev.} \\
\colhead{(km s$^{-1}$)} & \colhead{( mag)} 
& \colhead{($10^{-3}$ mag)} &
\colhead{$(10^{-22}$ mag cm$^{2}$)}
& \colhead{($10^{-3}$ mag)} 
} 
\startdata 
All & All & $1.7\pm0.8 $  & $1.0\pm0.03$& $3.5$ \\
\llap{$>-$}$100$ & All  &$-7.3\pm0.4 $ &$1.5\pm0.02$ & $2.0$ \\
\llap{$>-$}$100$ & \llap{$<$}$0.03$ &$-2.1\pm5.0 $ &$1.2\pm0.02$ & $1.6$ \\
\enddata 
\tablecomments{Error estimates are $\pm1\sigma$ from the 
least squares fit.}
\end{deluxetable}

\vfill \eject

\end{document}